\documentclass[prb,twocolumn,superscriptaddress,showpacs]{revtex4}
\usepackage{bm}
\usepackage{amsmath}
\usepackage{amssymb}
\usepackage[dvips]{graphicx}
\begin{document}
\title{Polaronic effects in electron shuttling}

\author{G. A. Skorobagatko}
\email{gleb_skor@mail.ru}
\affiliation{B. Verkin Institute for Low Temperature Physics and Engineering of
the National Academy of Sciences of Ukraine, 47 Lenin Avenue, Kharkov 61103, Ukraine}
\author{I. V. Krive}
\affiliation{B. Verkin Institute for Low Temperature Physics and Engineering of
the National Academy of Sciences of Ukraine, 47 Lenin Avenue, Kharkov 61103, Ukraine}
\affiliation{Department of Physics, University of Gothenburg, SE-412
96 G{\" o}teborg, Sweden}
\author{R. I. Shekhter}
\affiliation{Department of Physics, University of Gothenburg, SE-412
96 G{\" o}teborg, Sweden}

\date{\today}

\begin{abstract}
Shuttle-like mechanism of electron transport through a single
level vibrating quantum dot is considered in the regime of strong
electromechanical coupling. It is shown that the increment of
shuttle instability is a nonmonotonic function of the driving
voltage. The interplay of two oppositively acting effects -
vibron-assisted electron tunneling and polaronic blockade -
results in oscillations of the increment on the energy scale of
vibron energy.
\end{abstract}


\maketitle

\section{Introduction}

The modern trends in miniaturization of electronic devices
eventually led to fabrication of single molecular junctions and
molecular transistors (see e.g. review \cite{PR}). The electric
properties of single molecular transistors (SMTs) in many cases
are similar to the analogous characteristics of single electron
transistors (SETs) fabricated in two-dimensional electron gas.
SMTs demonstrate such effects as Coulomb blockade and Coulomb
blockade oscillations on gate voltage. The significant difference
between SMT and semiconducting SET is that the former can function
even at room temperatures that makes them to be very promising
basic elements for nanoelectronics.

Another specific feature of molecular transistors is the
interaction between electronic and vibrational degrees of freedom.
The electron in the process of tunneling through the molecule can
excite (and absorb at finite temperatures) molecular vibrational
quanta (vibrons) - the phenomenon known as "phonon-assisted
tunneling" \cite{GlSh}. The opening of inelastic channels results
in appearance of additional peaks (side-band peaks) in
differential conductance. For weak electron-vibron interaction the
magnitudes of inelastic peaks are much smaller then the value of
the elastic resonance peak. The situation is changed in the regime
of strong electron-vibron interaction when nonperturbative (and,
in particular, polaronic) effects determine electron transport
through a vibrating molecule (see e.g.\cite{PR}).

Polaronic effects are most pronounced in the case when the
molecule (quantum dot) is well-separated from the leads and the
width, $\Gamma$, of conducting molecular states is small compared
to other energy scales (temperature $T$, driving voltage $eV$). In
this case the mechanism of electron transport through the
vibrating molecule is (inelastic) sequential tunneling via the
real polaronic intermediate state. The amplitude of this tunneling
is exponentially suppressed since the wave functions of free
electron in the leads and polaronic state (Holstein polaron) in
the dot are almost orthogonal. This effect (named as Frank-Condon
\cite{Kch},\cite{Kcha} or polaronic \cite{Krive} blockade)
strongly suppresses elastic channel of electron transport and
changes the temperature behavior of conductance. Recently,
Frank-Condon blockade was observed in experiment of electron
tunneling through a suspended single-wall carbon nanotube
\cite{Leturcq}.

A one more novel phenomenon appears for vibrating quantum dot when
the matrix element of electron tunneling to the left and to the
right lead differently depends on the position of the dot center
of mass. This is always the case when the dot (molecule) vibrates
in the direction of electron tunneling and then the effect of
electron shuttling takes place at finite voltages
\cite{2.2}-\cite{2.4}. In papers \cite{2.3},\cite{2.4} the problem
of electron shuttling was considered for a model of single level
($\varepsilon_{0}$) vibrating quantum dot weakly coupled to the
leads of noninteracting electrons. It was shown that in the regime
of weak electromechanical coupling the shuttle instability occurs
at bias voltages $eV\geq 2(\varepsilon_{0}+\hbar \omega_{0})$
($\hbar \omega_{0}$ is the vibron energy) and the increment of
instability is $r_{0}\sim (\Gamma_{0}/\hbar)\lambda\lambda_{t}$.
Here $\Gamma_{0}$ is the level width, $\lambda$ is the
dimensionless electron-vibron interaction constant and
$\lambda_{t}=x_{0}/l_{t}$ ($x_{0}$ is the amplitude of zero-point
fluctuations of quantum dot, $l_{t}$ is the electron tunneling
length). Both coupling constants $\lambda$, $\lambda_{t}$ were
assumed to be small in Refs.\cite{2.3},\cite{2.4}.

In the problem of electron shuttling the electron-vibron
interaction strength $\lambda(V)$ linearly depends on the driving
voltage. Therefore, at sufficiently high voltages the regime of
strong electron-vibron interaction ($\lambda\gtrsim 1$) is
realized. In this regime polaronic effects could play significant
role in electron shuttling. In the present paper we reconsider the
problem of shuttle instability for the regime of strong
electromechanical coupling.

\begin{figure}
\includegraphics[height=7 cm,width=8.6 cm]{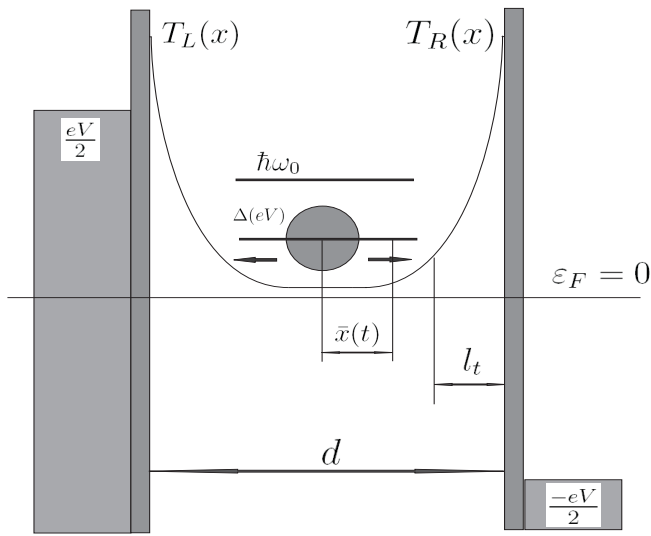}
\caption{Schematical diagram of Single Electron Transistor (SET) with
vibrating quantum dot (QD). Here $\Delta(eV)$ denotes the
"shifted" (due to polaronic shift) bias voltage-dependent fermionic level; $\hbar \omega_{0}$ is the energy of vibrational mode; $T_{L(R)}(x)$ are the coordinate-dependent tunneling amplitudes and $\pm eV/2$ are the chemical potentials of the leads $\varepsilon_{F}=0$. Characteristic distances in the QD: $d$ is the distance ("gap") between the leads; $l_{t}$ is the tunneling length of the electron in the QD; $\bar{x}(t)$ is average coordinate of the shuttle.}
\end{figure}

We derived the equation of motion for the shuttle average
coordinate $\bar{x}(t)$ assuming only the weak character of
dot-lead interaction. It was shown that the increment $r_{s}(V)$
of shuttle instability is a nonmonotonic function of bias voltage
with a maximum at $eV_{m}\sim \hbar \omega_{0}(d/x_{0})^{2}$,
where $d\gg x_{0}$ is the distance between the leads. The maximum
value of the increment is sensitive to value of coupling constant
$\lambda_{t}$. The interplay of two effects - the increase of
$r_{s}$ caused by the increase of inelastic channels which
contribute to the increment when rising the applied voltage and
the decrease of $r_{s}$ with the increase of bias voltage caused
by polaronic blockade - results in oscillation of $r_{s}$ on small
energy scale $\hbar \omega_{0}$.

Our results show that polaronic effects determine the physics of
electron shuttling in the case of moderate or strong mechanical
damping when the transition to a shuttle-like regime of electron
transport is possible only at sufficiently high bias voltages.

\section{The Model}

Our starting point is the model of vibrating quantum dot weakly
coupled to the leads of noninteracting electrons. This model was
repeatedly considered in the literature for the problem of electron
transport in molecular transistors (see e.g.\cite{PR} and
referencies therein). We expand this model to the problem of
electron shuttling \cite{2.2} by taking into account the explicit
dependence of tunneling amplitude on the center of mass coordinate
of quantum dot.

For simplicity we will study the case of a single level (with the
energy $\varepsilon_{0}$) quantum dot coupled to a single vibronic
mode (with the energy $ \hbar\omega_{0}$). The total Hamiltonian of
our system is
\begin{equation} \label{1}
\hat{H}=\sum_{j=L,R}\hat{H}_{l}^{(j)}+\hat{H}_{d}+\sum_{j=L,R}
\hat{H}_{t}^{(j)}
\end{equation}
where
\begin{equation} \label{2}
\hat{H}_{l}^{(j)}=\sum_{k}(\varepsilon_{k}-\mu_{j})\hat{a}_{kj}^{+}\hat{a}_{kj}
\end{equation}
is the standard Hamiltonian of noninteracting electrons
($\varepsilon_{k}$) in the left ($j=L$) and right ($j=R$) leads,
$\mu_{j}$ is the corresponding chemical potential:
$\mu_{L}-\mu_{R}=eV$ ( $V$ is the driving voltage);
$\hat{a}_{kj}^{+}$ ($\hat{a}_{kj}$) are the creation(destruction)
operators. The Hamiltonian of vibrating quantum dot takes the form
(see e.g.\cite{2.5})
\begin{equation} \label{3}
\hat{H}_{d}=\varepsilon_{0}\hat{c}^{+}\hat{c}-\varepsilon_{i}(\hat{b}^{+}+\hat{b})\hat{c}^{+}\hat{c}+\hbar\omega_{0}\hat{b}^{+}\hat{b}\,,
\end{equation}
where $\varepsilon_{i}$ is the characteristic energy of
electron-vibron interaction (see below), $\hat{c}^{+}$($\hat{c}$)
and $\hat{b}^{+}$($\hat{b}$) are fermionic and bosonic creation
(destruction) operators with commutation relations
$\{\hat{c},\hat{c}^{+}\}=1$; $[\hat{b},\hat{b}^{+}]=1$. The
tunneling Hamiltonian is
\begin{equation} \label{4}
\hat{H}_{t}^{(j)}=\sum_{k}\left[\hat{t}_{j}(\hat{X}_{c.m.})\hat{a}_{kj}^{+}\hat{c}+h.c.\right]\,.
\end{equation}
In Hamiltonian Eq.(\ref{4}) we take into account the dependence of
tunneling amplitude $t_{j}$ on the center of mass coordinate
$X_{c.m.}$ of quantum dot. In quantum description the coordinate
$X_{c.m.}$ becomes an operator
\begin{equation} \label{5}
\frac{\hat{X}_{c.m.}}{x_{0}}=\hat{X}=\frac{1}{\sqrt{2}}\left(\hat{b}^{+}+\hat{b}\right)\,;\
\hat{P}=\frac{i}{\sqrt{2}}\left(\hat{b}^{+}-\hat{b}\right)\,;
\end{equation}
where $x_{0}=\sqrt{\hbar/M\omega_0}$, ($M$ is the mass of QD) is
the amplitude of zero-point oscillations of quantum dot. In
Eq.(\ref{5}) we defined also the dimensionless momentum operator
$\hat{P}$ with the canonical commutation relation
$[\hat{X},\hat{P}]=i$. In what follows the tunneling amplitude is
model \cite{2.3} by the exponential function
\begin{equation} \label{6}
\hat{t}_{j}(\hat{X})=t_{0j}\exp(j\lambda_{t}\hat{X}),
\qquad j=(L,R)\equiv(-,+)\,.
\end{equation}
Here $\lambda_{t}\equiv x_{0}/l_{t}$ and $l_{t}$ is the tunneling
length.

Notice that electron-vibron interaction term in Eq.(\ref{3}) in our
model originates
from the electrostatic interaction of charge density on the dot
with the electrostatic potential produced by the leads \cite{2.3}.
It is convinient to
characterize this interaction by the dimensionless bias
voltage-dependent coupling constant
\begin{equation} \label{7}
\lambda=\frac{\sqrt{2}\varepsilon_{i}}{\hbar\omega_{0}}=r_{d}\frac{eV}{2\hbar\omega_{0}}\,,\qquad
r_{d}=\frac{2x_0}{d}\ll1
\end{equation}
($V$ is the bias voltage, $d$ is the distance between the leads).
We use the notations usually accepted in the literature on
molecular transistors. Notice that our notations for the coupling
constants differ from Refs.\cite{2.2,2.3,2.4}.

The problem of electron shuttling in the model Eqs.(\ref{1})-
(\ref{3}) was
studied in Refs.\cite{2.3,2.4} for the case of weak
electromechanical coupling (in our notations: $\lambda\ll1$,
$\lambda_{t}\ll1$). In molecular
transistors the electron-vibron interaction can be strong
$\lambda\gtrsim 1$ (see e.g. \cite{PR}). Here we reconsider
the problem of shuttle
instability \cite{2.3} in the regime of strong coupling.

To study electron transport in the presence of polaronic effects
($\lambda\gtrsim 1$) it is convenient to use unitary
transformation (see e.g.\cite{2.5}) which eliminates
electron-vibron interaction term in the dot Hamiltonian Eq.(\ref{3}).
Shuttle instability results in appearance of classical
time-dependent coordinate $\bar{x}(t)$ of quantum dot. It means
that bosonic operators $\hat{b}$ and $\hat{b}^{+}$ acquire
$c$-number part $\alpha(t)$:
\begin{equation} \label{8}
\hat{b}=\alpha(t)+\hat{\tilde{b}}\,;\
\hat{b}^{+}=\alpha^{*}(t)+\hat{\tilde{b}}^{+}\,,
\end{equation}
here $\hat{\tilde{b}}^{+}$($\hat{\tilde{b}}$) are the bosonic
creation(destruction) operators, which describe quantized vibron
modes
$\langle\hat{\tilde{b}}\rangle=\langle\hat{\tilde{b}}^{+}\rangle=0$
($\langle\ldots\rangle$ denotes the thermal average).

We transform total Hamiltonian (1) using the unitary operator (see
e.g.\cite{PR})
\begin{equation} \label{9}
\hat{U}=\exp(i\lambda\hat{p}\hat{n})\,,\,\hat{n}=\hat{c}^{+}\hat{c}\,,\,\hat{p}=\frac{i}{\sqrt{2}}\left(\hat{\tilde{b}}^{+}-\hat{\tilde{b}}\right)\,.
\end{equation}
The unitary transformed Hamiltonian takes the form
\begin{equation} \label{10}
\hat{\tilde{H}}=\hat{U}\hat{H}\hat{U}^{-1}=\varepsilon(t)\hat{c}^{+}\hat{c}+\hbar\omega_{0}\hat{b}^{+}\hat{b}+\sum_{j=L,R}\hat{\tilde{H}}_{t}^{(j)}+\sum_{j=L,R}\hat{\tilde{H}}_{l}^{(j)}
\end{equation}
where
$\varepsilon(t)=\varepsilon_{0}-[\lambda\bar{x}(t)+\lambda^{2}/2]\hbar\omega_{0}$,
$\bar{x}(t)=\sqrt{2}Re\alpha(t)$. The time-independent shift
($\sim\lambda^{2}$) of the dot level is called polaronic shift
\cite{PR}. The transformed tunneling Hamiltonian in Eq.(\ref{10}) is
\begin{eqnarray}
\label{11}
\hat{\tilde{H}}_{t}^{(j)}=\sum_{k}\left[\hat{t}_{j}(\hat{X}_{t})\hat{a}_{kj}^{+}e^{-i\lambda\hat{p}}\hat{c}+h.c.\right]\nonumber\\
=\bar{t}_{0j}\sum_{k}\left[\hat{t}_{j}(\hat{X}_{t})\hat{a}_{kj}^{+}\hat{V}_{j}^{+}\hat{c}+h.c.\right]\,.
\end{eqnarray} 
Here $\hat{X}_{t}$ is the transformed coordinate operator
$\hat{X}_{t}=\hat{U}\hat{X}\hat{U}^{-1}=\hat{X}+\lambda\hat{n}$
and
\begin{eqnarray}
 \label{12}
\hat{V}_{j}=\exp(j\lambda_{t}\hat{X}+i\lambda\hat{p})\,,\nonumber\\
\,\bar{t}_{0j}=t_{0j}\exp(j\frac{\lambda\lambda_{t}}{2})\,,\,j=(L,R)\equiv(-,+)\,.
\end{eqnarray}

\section{Equations of motion}

The Heisenberg equations of motion for dimensionless operators of
coordinate $\hat{X}$ and momentum $\hat{P}$
\begin{eqnarray}
 \label{13}
\hat{X}=\frac{1}{\sqrt{2}}\left(\hat{b}^{+}+\hat{b}\right)=\bar{x}(t)+\hat{x}\,,\nonumber\\
\qquad\hat{P}=\frac{i}{\sqrt{2}}\left(\hat{b}^{+}-\hat{b}\right)=\bar{p}(t)+\hat{p}\,.
\end{eqnarray}
(where $\bar{x}(t)=\sqrt{2}Re\alpha(t)$,
$\bar{p}(t)=\sqrt{2}Im\alpha(t)$ ) can be represented in the form
of Hamilton equations
\begin{equation} \label{14}
\frac{d\hat{X}}{dt}=\frac{1}{\hbar}\frac{\partial\hat{H}}{\partial\hat{P}}\,,\,\,\,\,\frac{d\hat{P}}{dt}=-\frac{1}{\hbar}\frac{\partial\hat{H}}{\partial\hat{X}}\,,
\end{equation}
with the Hamiltonian $\hat{H}$ given by the following expression
\begin{equation} \label{15}
\hat{H}=\frac{\hbar\omega_{0}}{2}\left(\hat{X}^{2}+\hat{P}^{2}\right)+\sum_{j=L,R}\hat{H}_{j}(\hat{X},\hat{p})\,.
\end{equation}
Here we denote by $\hat{H}_{j}$ the tunneling Hamiltonian defined
by Eq.(\ref{11}). In our model equations (\ref{14}) take the form
\begin{eqnarray}
 \label{16}
\frac{d\hat{X}}{dt}=\omega_{0}\hat{P}-\lambda\sum_{j=L,R}\hat{J}_{j}(\hat{X},\hat{p})\,,\nonumber\\
\qquad\frac{d\hat{P}}{dt}=-\omega_{0}\hat{X}-\frac{\lambda_{t}}{\hbar}\sum_{j=+/R,-/L}j\hat{H}_{j}(\hat{X},\hat{p})\,,
\end{eqnarray}
where $\hat{J}_{L,R}$ are the current operators
\begin{equation} \label{17}
\hat{J}_{j}=i\frac{\bar{t}_{0j}}{\hbar}\sum_{k}\left[\hat{a}_{kj}^{+}\hat{V}_{j}^{+}\hat{c}-h.c.\right]\,.
\end{equation}
These operators satisfy (as it should be) the continuity equation
\begin{equation} \label{18}
\frac{d\hat{n}}{dt}=\hat{J}_{L}+\hat{J}_{R}\,.
\end{equation}
With the help of Eq.(\ref{18}) we can rewrite the first expression of
Eq.(\ref{16}) in the following form
\begin{equation} \label{19}
\hat{P}=\omega_{0}^{-1}\frac{d}{dt}\left(\hat{X}+\lambda\hat{c}^{+}\hat{c}\right)\,,
\end{equation}
and the second equation in (\ref{16}) is transformed to
\begin{equation} \label{20}
\frac{d^{2}}{dt^{2}}\left(\hat{X}+\lambda\hat{c}^{+}\hat{c}\right)+\omega_{0}^{2}\hat{X}=-\frac{\omega_{0}\lambda_{t}}{\hbar}
\sum_{j=-/L,+/R}j\hat{H}_{j}(\hat{X},\hat{p})\,,
\end{equation}
where $\hat{p}=\hat{P}-\bar{p}(t)$.

To make the operator equations (Eq.(\ref{16}) or Eqs.(\ref{19})),
(\ref{20}) complete
we have to write down the equations of motion for fermionic operators
$\hat{a}_{kj}$($\hat{a}_{kj}^{+}$) and $\hat{c}$($\hat{c}^{+}$).
These equations
\begin{equation} \label{21}
\frac{d\hat{a}_{kj}}{dt}=-i(\varepsilon_{k}-\mu_{j})\hat{a}_{kj}(t)-i\bar{t}_{0j}\hat{V}_{j}^{+}(t)\hat{c}(t)\,,
\end{equation}
\begin{equation} \label{22}
\frac{d\hat{c}}{dt}=-i\varepsilon(t)\hat{c}(t)-i\sum_{k,j=L,R}\bar{t}_{0j}\hat{V}_{j}(t)\hat{a}_{kj}(t)\,
\end{equation}
are linear and they can be readily solved (the equations of motion
for creation operators are obtained from Eqs.(\ref{21}),(\ref{22})
simply by taking Hermitian conjugation). We will follow
Refs.\cite{2.3} and find the solution of Eqs.(\ref{21}),(\ref{22})
in the so called ''wide band approximation"("WBA")(see e.g.
\cite{WBA}). The only difference of our system of equations
(\ref{21}),(\ref{22}) from the corresponding one in Ref.\cite{2.3}
is the presence in our case bosonic operator factors $\hat{V}_{j}$
and $\hat{V}_{j}^{+}$. Formally, these factors make the level
width $\hat{\Gamma}$ (see the definition below in Eq.(\ref{24}), which
appears in Eq.(\ref{22}) after substituting in this equation the
solution of Eq.(\ref{21}), to be an operator as well. The
dynamical equation for the dot level operator $\hat{c}$ in WBA
takes a simple form
\begin{eqnarray}
 \label{23}
\frac{d\hat{c}}{dt}=-i\left[\varepsilon(t)-i\frac{\hat{\Gamma}(t)}{2}\right]\hat{c}(t)\nonumber\\
-i\sum_{k,j=L,R}\bar{t}_{0j}\hat{V}_{j}(t)\hat{a}_{kj}(0)e^{-i(\varepsilon_{k}-\mu_{j})t}\,,
\end{eqnarray}
where
\begin{equation} \label{24}
\frac{1}{2}\hat{\Gamma}(t)=\sum_{j=L,R}\Gamma_{0j}\hat{V}_{j}(t)\hat{V}_{j}^{+}(t)\,.
\end{equation}
Here $\Gamma_{0j}=2\pi\rho_{j}|\bar{t}_{0j}|^{2}$ and
$\rho_{j}$ is the density of states in the $j=L,R$
electrode ($\rho_{j}$ is assumed to be energy independent in the
wide band approximation).

To proceed further we have to make additional simplifications. Our
purpose in this section is to derive the equation of motion for
the average coordinate in the presence of fluctuations. Notice
that in the Hamiltonian Eq.(\ref{10}) electron-vibron interaction
appears only in the tunneling term which for a weak tunneling can
be treated perturbatively. In perturbation theory on the bare
level widths $\Gamma_{0j}$ vibrons are decoupled from fermions and
when evaluating averages of the products of fermion ($\hat{F}$)
and boson ($\hat{B}$) operators we can use the decoupling
procedure
$\,\langle\hat{F}\hat{B}\rangle\backsimeq\langle\hat{F}
\rangle_{\tilde{H}_{f}+\tilde{H}_{t}}\langle\hat{B}\rangle_{\tilde{H}_{b}}\,.$
The averages of fermion operators are taken with "fermionic"
part of the total transformed Hamiltonian (including the
tunneling part) and averages of boson operators are
calculated with the quadratic Hamiltonian of
nonointeracting vibrons
$\hat{\tilde{H}}_{b}=\hbar\omega_{0}\hat{\tilde{b}}^{+}\hat{\tilde{b}}$.
This decoupling procedure is the starting point of calculations in
many papers dealing with the polaronic effects in electron
transport in molecular junctions (see e.g. review \cite{PR} and
referencies therein).

The linear equation (\ref{23}) can be formally solved (using
time-ordering procedure) for the operator level
width $\hat{\Gamma}(t)$.
However the closed equation for the average coordinate
$\bar{x}(t)$ can be obtained only in perturbation theory on
$\Gamma_{0j}$. In Ref.\cite{2.4} such equation was obtained in the
limit of weak electromechanical coupling $\lambda\ll1$,
$\lambda_{t}\ll1$. Here we derive the equation for classical
coordinate $\bar{x}(t)$ valid also for the regime of strong
electron-vibron coupling $\lambda\gtrsim1$.

In perturbation theory on the bare level width $\Gamma_{0j}$ we
can neglect the time dynamics of the level width and replace
$\hat{\Gamma}(t)$ by constant
$\Gamma_{0}=\Gamma_{0L}+\Gamma_{0R}$. It is convinient in what
follows to represent "vertex" operator $\hat{V}_{j}$ as a product
of classical $T_{j}\{\bar{x}(t)\}$ and quantum $\hat{Q}_{j}$ parts
$\hat{V}_{j}=T_{j}(t)\hat{Q}_{j}$

\begin{eqnarray}\label{25}
\hat{T}_{j}(t)=\exp[j\lambda_{t}\bar{x}(t)]\,,\,\hat{Q}_{j}(\hat{x},\hat{p})=\exp(j\lambda_{t}\hat{x}+i\lambda\hat{p})\,,\nonumber\\
\,j=(L,R)\equiv(-,+)\,
\nonumber\\
\end{eqnarray}
In the discussed approximation the solution of Eq.(\ref{23}) takes the
form
\begin{eqnarray}\label{26}
\hat{c}(t)=-i\sum_{k,j=L,R}\bar{t}_{0j}\hat{a}_{kj}(0)e^{-i(\varepsilon_{k}-\mu_{j})t}\nonumber\\
 \cdot \int_{0}^{t}dt_{1}T_{j}(t)\hat{Q}_{j}\exp\left\{i\int_{t_{1}}^{t}dt_{2}\left[\varepsilon_{k}-\mu_{j}-\varepsilon(t_{2})+\frac{i}{2}\Gamma_{0}\right]\right\}\,.
\nonumber\\ 
\end{eqnarray}

The equation of motion for the classical coordinate $\bar{x}(t)$
in perturbation theory on $\Gamma_{0j}$ can be readily obtained
from the exact operator equation (\ref{20}) by taking the average
and using the discussed above "fermion-boson" factorization
procedure
\begin{equation} \label{27}
\frac{d^{2}}{dt^{2}}\left[\bar{x}(t)+\lambda
N\{\bar{x}(t)\}\right]+\omega_{0}^{2}\bar{x}(t)=-\frac{\omega_{0}\lambda_{t}}{\hbar}\sum_{j=_{+/R}^{-/L}}jH_{j}\left\{\bar{x}(t)\right\}\,,
\end{equation}
where
\begin{equation} \label{28}
N\{\bar{x}(t)\}=\langle\hat{c}^{+}(t)\hat{c}(t)\rangle\,,\qquad
H_{j}\left\{\bar{x}(t)\right\}=\langle\hat{H}_{j}(\hat{X},\hat{p})\rangle\,
\nonumber\\
\end{equation}
It is useful to rewrite Eq.(\ref{27}) in terms of a new variable
$\bar{x}_{v}(t)=\bar{x}(t)+\lambda N\{\bar{x}(t)\}$. Notice that
according to Eqs.(\ref{11}),(\ref{26}) both quantities in
Eq.(\ref{27}) - averaged tunneling Hamiltonian $H_{j}$ and the
level occupation number $N\{\bar{x}(t)\}$ - are proportional to
the bare level width  $H_{j},N\{\bar{x}(t)\}\propto\Gamma_{0j}$.
Since Eq.(\ref{27}) is derived in the Born approximation (up to
the second order in the tunneling amplitude) we can replace
$\bar{x}(t)$ by $\bar{x}_{v}(t)$ in the averaged quantities. Then
Eq.(\ref{27}) for the variable $\bar{x}_{v}(t)$ (shifted
coordinate) takes the form of the Newton's equation derived in
Ref.\cite{2.4}
\begin{equation} \label{29}
\frac{d^{2}}{dt^{2}}\bar{x}_{v}(t)+\omega_{0}^{2}\bar{x}_{v}(t)=\bar{F}_{q}\{\bar{x}_{v}(t)\}\,,
\end{equation}
where
\begin{equation} \label{30}
\bar{F}_{q}\{\bar{x}(t)\}=\lambda\omega_{0}^{2}N\{\bar{x}(t)\}-\frac{\omega_{0}\lambda_{t}}{\hbar}\sum_{j=_{+/R}^{-/L}}jH_{j}\left\{\bar{x}(t)\right\}\,
\end{equation}
This equation in the limit $\lambda\ll1$, $\lambda_{t}\ll1$, when
we can omit operator factors ($\hat{Q}_{j}$) in the vertex
function $\hat{V}_{j}$, exactly coincides with the corresponding
equation for the coordinate of ''classical" shuttle (see
Refs.\cite{2.3,2.4}). One can analyze shuttle instability  by
using either Eq.(\ref{27}) or Eq.(\ref{29}). The only difference
is the starting (equilibrium) position of the shuttle. This is
$\bar{x}=0$ for Eq.(\ref{27}) and $\,\bar{x}_{v}=\lambda N\neq 0$
for Eq.(\ref{29}) (see Ref.\cite{2.4}). We will use Eq.(\ref{27}).

\section{The increment of shuttle instability}

The conditions for a shuttle instability can be found by analyzing
linearized equation of motion. We will follow Ref.\cite{2.4} where
these conditions were obtained for the regime of weak
electromechanical coupling. The linear integral-differential
equation for the shuttle coordinate $\bar{x}(t)$ in the
dimensionless units ($t\rightarrow \omega_0t$, energy scale is
$\hbar\omega_0$) reads (see Appendix)
\begin{eqnarray}
 \label{31}
\frac{d^{2}}{dt^{2}}\bar{x}(t)+\bar{x}(t)=-\sum_{j=_{+/R}^{-/L}}\frac{\bar{\Gamma}_{0j}}{2\pi}\sum_{l=-\infty}^{+\infty}F_{lj}(\beta)\nonumber\\
\cdot \int_{-\infty}^{+\infty}d\varepsilon f_{j}(\varepsilon)\left\{2Im\left[X_{lj}(\varepsilon,t)\right]\left|\lambda
G_{lj}^{(+)}(\varepsilon)-j\lambda_{t}\right|^{2}\right\}\,
\nonumber\\
\end{eqnarray}
In Eq.(\ref{31}) the following notations are introduced:
$\,\bar{\Gamma}_{0j}=\Gamma_{0j}\exp(j\lambda\lambda_{t})$
\begin{eqnarray}
 \label{32}
X_{lj}(\varepsilon,t)=\int_{0}^{t}d\tau
\bar{x}(\tau)\nonumber\\
\cdot \exp\left\{i\left[(\varepsilon-\mu_{j})-(\Delta-l)+
i\frac{\Gamma_{0}}{2}\right](t-\tau)\right\}\,,
\end{eqnarray}
and
\begin{equation} \label{33}
G_{lj}^{(\pm)}(\varepsilon)=\frac{1}{(\varepsilon-
\mu_{j})-(\Delta-l)\pm
i\Gamma_{0}/2}\,
\end{equation}
where $\Delta=\varepsilon_{0}-\lambda^{2}/2\,$ is the
polaronic shift. Vibrational degrees of freedom result
in Franck-Condon factors
\begin{eqnarray}
 \label{34}
F_{lj}(\beta)=
\exp[-(\lambda^{2}-\lambda_{t}^{2})(1+2n_{b})]
\left|\frac{\lambda+j\lambda_{t}}{\lambda-j\lambda_{t}}\right|^{l}\nonumber\\
\cdot I_{l}(2|\lambda^{2}-\lambda_{t}^{2}|\sqrt{n_{b}(1+n_{b})})e^{-\beta
l/2}\,
\end{eqnarray}
Here $I_{l}(x)$ is the modified Bessel function of the
second kind and
$n_{b}=1/(\exp\beta-1)\,$, ($\beta=\hbar\omega_0/k_BT\,$),
$f_{j}(\varepsilon)=1/[\exp(\beta(\varepsilon-\mu_{j}))+1]$
are Bose-Eistein and Fermi-Dirac
distribution functions. Remind that all energies
in above expressions are dimensionless
(in the units of $\hbar\omega_0$). In the limit $\lambda_t=0$
the vibron-induced correlation factor Eq.(\ref{34}) coincides
with the well-known in the literature expression (see e.g.
Refs.(\cite{2.5,LMcK}).

The solution of Eq.(\ref{31}) with the initial condition
$\bar{x}(0)=0$ is
\begin{equation} \label{35}
\bar{x}(t)=A_{0}e^{r_{s}t}\sin(t)\,.
\end{equation}
where $A_0$ is an arbitrary constant and
\begin{eqnarray}
\label{36}
r_{s}=\sum_{j=_{+/R}^{-/L}}\frac{\bar{\Gamma}_{0j}}{8}\sum_{l=-\infty}^{+\infty}F_{lj}(\beta)(\lambda-j\lambda_{t})^{2}f_{j}(\Delta-l+1)\nonumber\\
-\sum_{j=_{+/R}^{-/L}}\frac{\bar{\Gamma}_{0j}}{8}\sum_{l=-\infty}^{+\infty}F_{lj}(\beta)(\lambda+j\lambda_{t})^{2}f_{j}(\Delta-l-1)\,
\nonumber\\
\end{eqnarray}
is the dimensionless increment (when $r_{s}>0$) of shuttle
instability. Our purpose here is to find conditions (driving
voltage) for the realization of shuttle motion in the presence
of strong electron-vibron interaction.

For a weak electromechanical coupling $\lambda\ll1$\,,
$\lambda_{t}\ll1$ we can neglect quantum and
thermodynamical fluctuations of vibrons
(they are of higher orders on coupling constants) and omit
all terms in the right-hand side of Eq.(\ref{36}) but $l=0$.
In this limit $\Delta\simeq\varepsilon_{0}$,
$\bar{\Gamma}_{0j}\simeq\Gamma_{0j}$, $F_{0j}\simeq 1$ and
Eq.(\ref{36}) is transformed to the corresponding equation
of Refs.\cite{2.3,2.4}. The increment Eq.(\ref{36}) in this
case coincides with the one found earlier Ref.\cite{2.4}
\begin{eqnarray}
\label{37}
r_{s}^{(0)}=\sum_{j=_{+/R}^{-/L}}\frac{\bar{\Gamma}_{0j}}{8}(\lambda-j\lambda_{t})^{2}f_{j}(\varepsilon_{0}+1)\nonumber\\
-\sum_{j=_{+/R}^{-/L}}\frac{\bar{\Gamma}_{0j}}{8}(\lambda+j\lambda_{t})^{2}f_{j}(\varepsilon_{0}-1)\,
\end{eqnarray}
(there is a misprint in Eq.(24) of Ref.\cite{2.4} - the
sign-changing factor $j$ is missing in the coefficients
in front of distrubution functions).

It is qualitatively clear that shuttle instability is most
pronounced at low temperatures $\beta\gg1$ when there are no
thermally activated vibrons. At low temperatures shuttle
instability results in a sharp (step-like) features in the $I-V$
curve of single electron transistor \cite{2.3}. Finite temperature
effects broaden the transition region and at $T\gg\hbar\omega_0$
the shuttle instability is less pronounced (nevertheless as far as
$r_s>0$ the instability results in a strong increase of electric
current). In what follows we will consider only low-$T$ effects
and (for simplicity) the case of a symmetric junction
($\Gamma_{0L}=\Gamma_{0R}=\Gamma\,,\mu_{L}=-\mu_{R}=eV/2 $, we set
Fermi energy of the leads $\varepsilon_F=0 $). At low temperatures
the $l>0$ terms in Eq.(\ref{36}) are exponentially suppressed in
comparison with the negative $l$ (formally, due to the factor
$\exp(-l\beta/2)$ since $I_{-l}(x)=I_l(x)$ for the integer $l$).
Physically, positive $l$ corresponds to vibron absorption - an
energetically forbidden process at $T\ll\hbar\omega_0$. Negative
$l$ describes emission of vibrons and the summation over $l$ at
finite bias voltage is limited by a certain $l_m$ (see below).

By using the well-known asymptotics of the Bessel function at
small arguments $I_l(x\rightarrow 0)\sim (z/2)^l/l!$ and
replacing Fermi distribution functions in Eq.(\ref{36}) by
the Heaviside theta-functions we obtain the desired formula for
the increment ($r_s>0$) of shuttle instability
\begin{eqnarray}
 \label{38}
r_{s}=\frac{\Gamma\lambda\lambda_{t}}{2}
\exp(-\lambda^2+\lambda_t^2-\lambda\lambda_{t})\sum_{l=0}^{l_{m}-1}\frac{(\lambda+\lambda_{t})^{2l}}{l!}\nonumber\\
= \frac{\Gamma\lambda\lambda_{t}}{2(l_m-1)!}\exp(\lambda\lambda_t
+2\lambda_t^2)\Gamma\left(l_m,(\lambda+\lambda_t)^2\right)\,,
\end{eqnarray}
where $\Gamma(\alpha,x)$ is the incomplete gamma function (see
e.g. \cite{GR}) and
\begin{equation} \label{lm}
l_m=\left[\frac{eV}{2}-\left(\varepsilon_0-\frac{\lambda^2}{2}\right)-1\right]
\end{equation}
($[x]$ denotes the integer part of $x$). We find from
Eqs.(\ref{38}),(\ref{lm}) that in the regime of weak
electromechanical coupling the instability occurs at
$eV>eV_c\simeq 2(\varepsilon_0+1)$ \cite{2.3},\cite{2.4} and the
increment $r_s(\lambda\ll1, \lambda_t\ll1)\simeq
r_0=\Gamma\lambda\lambda_{t}/2\,$ (see Ref.\cite{2.4}) is a linear
function of bias voltage. Remind that we consider the case when
the uncertainty in the quantum dot initial position due to quantum
fluctuations ($\sim x_0$) is small compared to the geometrical
size $d$ of the junction (that is our parameter $r_d=2x_0/d\ll1$,
Eq.(\ref{7})). In this case at the threshold voltage $V_c$ the
electron-vibron coupling is small
$\lambda(V_c)=r_d(\varepsilon_0+1)\ll1$ (we always can put
$\varepsilon_0=0$ at the resonance condition) and polaronic
effects are not pronounced. Nevertheless there is a small negative
correction to the threshold voltage due to polaronic shift
$V_c=2(\varepsilon_0-\lambda^2(V_c)/2 +1)$ and the multiplicative
renormalization of the bare level width by quantum fluctuations
$\,\Gamma\rightarrow\Gamma\exp(-\lambda^2+
\lambda_t^2-\lambda\lambda_t)\simeq\Gamma(1-\lambda^2-\lambda
\lambda_t+\lambda_t^2)\,$, as one can see from
Eqs.(\ref{38})(\ref{lm}).

The increment $r_s$ is described by Eq.(\ref{38}) at bias voltages
in the interval $eV_c/2 < eV/2 \leq 2r_d^{-2}$. At higher voltages
the electron distribution function of the right electrode (biased
by $-eV/2$) in Eq.(\ref{36}) $f_R=1$ and the processes of
inelastic electron tunneling to the right bank start to contribute
to $r_s$. Their contributions according to Eq.(\ref{36}) at low
temperatures are negative and they could only diminish the
increment. We show now that due to polaronic (Franck-Condon)
blockade both the "right" and "left" contributions at voltages
$eV>2r_d^{-2}\gg 1$ are exponentially suppressed and shuttle
instability takes place in the finite interval of bias volages (we
restore here the dimensions)
\begin{equation} \label{biasV}
2(\varepsilon_0+\hbar\omega_0) < eV\leq 4\hbar\omega_0/r_d^2 \,
\end{equation}

The finite series on $l$ in Eq.(\ref{38}) can be approximated
as follows
\begin{equation} \label{S_l}
S_l(x) = \sum_{n=0}^l\frac{x^n}{n!}\simeq
\begin{cases} e^x,\qquad&l\gg x;\cr
              x^l/l!    &x\gg l\gg1. \cr
\end{cases}
\end{equation}
With the help of asymptotics Eq.(\ref{S_l})
we obtain the following formula
for $r_s$ at $eV\ll r_d^{-2}$ (the corresponding electron-vibron
coupling ''constant" $\lambda=r_d(eV/2)\ll r_d^{-1}$)
\begin{equation} \label{r_s}
r_s\simeq \frac{\Gamma\lambda\lambda_{t}}{2}
\exp(\lambda\lambda_t+2\lambda_t^2) \,
\end{equation}
For a very high biases $eV\gg r_d^{-2}$ it is easy to show from
Eqs.(\ref{38}),(\ref{S_l}) (using Sterling's formula to estimate
asymptotics of $l_m!$) that $\,r_s\propto\exp[-\lambda^2
(1-\ln2)/2]\rightarrow 0$ at $\lambda\gg1$. So we see that
the dependence of increment $r_s$ on the bias voltage is
strongly nonmonotonic with the maximum at $eV\sim r_d^{-2}$.

It is interesting to notice here that at the excitation energy
$E_{d}\simeq \hbar\omega_0(r_{d})^{-2}$ (the corresponding number
of excited vibrons $l_{d}\simeq r_{d}^{-2}\gg 1$) the
characteristic width, $w$, of the wave function of harmonic
oscillator which represents quantum dot in our model (Eq.(\ref{3}))
is of the order of gap, $d$, between the leads ($w\sim
x_{0}\sqrt{\langle \hat{x}^{2}\rangle}\sim x_{0}\sqrt{l_{d}}\sim
d$).

The approximation Eq.(\ref{S_l}) does not reveal the fine
structure (on the $\hbar\omega_0$ scale) of the dependence
$r_s(V)$. On this small energy scale one could expect the
appearance of steps each time the additional vibration channel
contributes to the increment (see Eq.(\ref{lm})). At low voltages
the steps are slightly modified by the Franck-Condon factors. With
the increase of voltage in the regime of strong coupling $\lambda
>1$ each additional channel modifies $r_s$ by the value of
the order (few times smaller) of $r_s$ (sharp big steps). However
between two sequential steps $r_s$ is diminished (due to the
$\exp(-\lambda^2(V))$ factor in Eq.(\ref{38})) approximately by
the same amount. That is on the scale of vibron energy there are
oscillations of $r_s$. These oscillations are smeared out at
temperatures $T\gg \hbar\omega_0$.

\begin{figure}
\includegraphics[height=6 cm,width=8.6 cm]{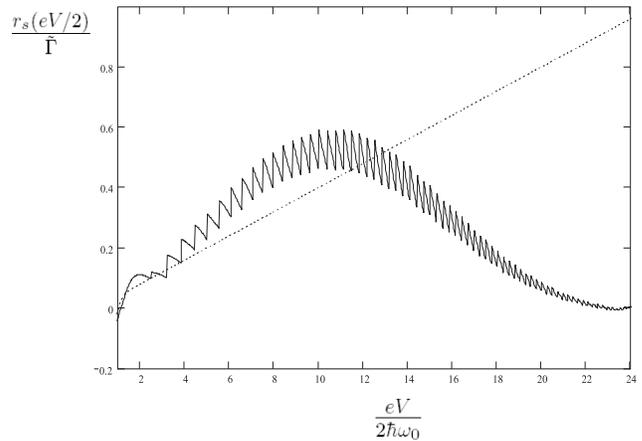}
\caption{"Weak" shuttle instability. The increment of shuttle instability (in the units of
$\Gamma/\hbar \omega_{0}$) as a function of bias voltage for
$\lambda_{t}=x_{0}/l_{t}=0.2$; $r_{d}=x_{0}/d=0.2$ (solid line)
and $\beta^{-1}=k_{B}T/\hbar\omega_{0}=0.2$. The dotted line
represents the result of Ref \cite{2.4} extended to the region of
strong electromechanical coupling.}
\end{figure}

\begin{figure}
\includegraphics[height=5.5 cm,width=8.6 cm]{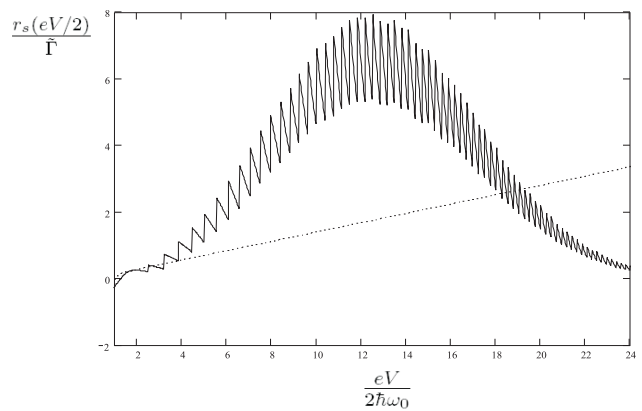}
\caption{"Strong" shuttle instability. All parameters are the
same as in Fig.2, but for the coupling constant $\lambda_{t}=0.7$.}
\end{figure}

The dependence $r_s(V)$ is plotted in Figs.2,3 for the different
values of parameters of our model (for numerical calculations we
used the basic Eq.(\ref{38}) and put the inverse temperature
$\beta=10$). For both figures we set $r_{d}=0.2$, since the
further decrease of $r_{d}$ yields the results which practically
coincide with Ref.\cite{2.4} in the considering region of applied
voltages. The dimensionless increment $r_{s}$ of the shuttle
instability is measured in the units of dimensionless
$\tilde{\Gamma}=\Gamma/\hbar\omega_{0}$ ($eV$ is in the units of
$\hbar\omega_{0}$). In Fig.2 we set $\lambda_{t}=r_{d}=0.2$, that
corresponds to the case when $2l_{t}\sim d$. In Fig.3 we put
$\lambda_{t}=0.7>r_{d}$, that corresponds to the case when
$2l_{t}< d$. The straight line on both figures shows the increment
for the "classical" shuttle of Ref.\cite{2.4} for the same values
of all parameters.

We see from the figures, that, the dependence of increment on bias
voltage is strongly nonmonotonic. The increment $r_{s}$ grows
until $eV\leq eV_{m}\approx r_{d}^{-2}$ and decreases when
$V>V_{m}$. The magnitudes of $r_{s}$ in Fig.3 ($\lambda_{t}=0.7$)
are much greater than the ones in Fig.2 ($\lambda_{t}=0.2$). This
means, that shuttle instability is very sensitive to the value of
tunneling lengths $\lambda_{t}$. So, when $l_{t}\ll d$ ($r_{d}\ll
1$), the greater is $\lambda_{t}$ the stronger is shuttle
instability (i.e. it develops on a shorter time scale $\sim
r_{s}^{-1}$). In the presence of strong mechanical friction,
characterized by phenomenological damping term,
$i\gamma_{f}\dot{\bar{x}}(t)$, introduced in Eq.(\ref{27}), and
for $2l_{t}\lesssim d$ (i.e. $r_{d}\lesssim 1$), when
$\lambda_{t}\ll 1$, the increment at all voltages could be less
then the friction coefficient $r_s<\gamma_{f}$. In this case
shuttle regime of electron transport is not realized.

\section{Summary}

In this paper we have studied the influence of quantum and
thermodynamical fluctuations on shuttle instability. These
fluctuations are significant in the case of strong
electromechanical coupling that can be realized in electron
transport through vibrating quantum dot at high bias voltages.

It was shown that the increment of shuttle instability is a
nonmonotonic function of bias voltage $V$ with a maximum at
$eV_{m}\sim \hbar \omega_{0}(d/x_{0})^{2}$, which corresponds to
the region of strong electron-vibron coupling $\lambda(V_{m})\gg
1$. At higher voltages polaronic blockade suppresses shuttle
instability. The maximum value of increment is sensitive to the
electromechanical dimensionless coupling constant
$\lambda_{t}=x_{0}/l_{t}$ ($l_{t}$ is the electron tunneling
length). In the presence of mechanical friction (characterized by
phenomenological friction coefficient $\gamma_{f}$) and when
$\lambda_{t}\ll 1$ the increment $r_{s}(V)$ at all voltages could
be less then friction $r_{s}(V)<\gamma_{f}$ and shuttle regime of
electron transport is not realized. In the regime of strong
electromechanical coupling $\lambda_{t}\sim 1$, $\lambda\gg 1$ the
pumping of energy in mechanical shuttle motion could overcome even
strong damping.

We showed that in this regime the increment of shuttle instability
strongly oscillates on the energy scale of vibron energy $\hbar
\omega_{0}$. If the friction coefficient is comparable with the
amplitude of increment oscillations, the small change of bias
voltage ($\Delta V \lesssim \hbar \omega_{0}/e$) drastically
changes the regime of electron transport through a vibrating
quantum dot (from a phonon-assisted tunneling to a shuttle regime
of electron transfer and vice-a-versa). One can speculate that
reentrant transitions to a shuttle-like regime of electron
transport will result in  unusual current-voltage characteristics
with pronounced negative differential conductance.

\section{Acknowledgments}

The authors thank S.I.Kulinich for valuable discussions. This work
was supported in parts by the Swedish VR and SSF, by the Faculty
of Science at the University of Gothenburg through the
"Nanoparticle" Research Platform and by the grant "Effects of
electronic, magnetic and elastic properties in strongly
inhomogeneous nanostructures" provided by the National Academy of
Sciences of Ukraine. IVK thanks the Department of Physics at the
University of Gothenburg for hospitality.

\section{The Appendix}

To study the shuttle instability we linearize the Eq.(\ref{29})
($\lambda\bar{x}(t)\ll1$, $\lambda_{t}\bar{x}(t)\ll1$). In the
linear approximation we have
$T_{j}(\bar{x}(t))=\bar{t}_{0j}(1+j\lambda_{t}\bar{x}(t))$,
$j=-/L;+/R$ and
$\hat{c}(t)=\hat{c}_{0}(t)+\hat{c}_{1}(t,\bar{x}(t))$, where
\begin{eqnarray}
\label{A3}
\hat{c}_{0}(t)=-i\sum_{k,j=L,R}\bar{t}_{0j}\hat{a}_{kj}(0)e^{-i(\varepsilon_{k}-\mu_{j})t}\nonumber\\
\cdot \int_{0}^{t}dt_{1}\hat{Q}_{j}(t_{1})\exp\left\{i\left[\varepsilon_{k}-\mu_{j}-\Delta+\frac{i}{2}\Gamma_{0}\right](t-t_{1})\right\}\,
\nonumber\\
\end{eqnarray}
and
\begin{eqnarray} \label{A4}
 \hat{c}_{1}(t)=-i\sum_{k,j=-/L,+/R}\bar{t}_{0j}\hat{a}_{kj}(0)e^{-i(\varepsilon_{k}-\mu_{j})t}\nonumber\\
 \cdot \int_{0}^{t}dt_{1}\left[j\lambda_{t}\bar{x}(t_{1})+\lambda\int_{t_{1}}^{t}dt_{2}\bar{x}(t_{2})\right]\nonumber\\
\cdot \hat{Q}_{j}(t_{1})\exp\left\{i\left[\varepsilon_{k}-\mu_{j}-\Delta+\frac{i}{2}\Gamma_{0}\right](t-t_{1})\right\}\,
\nonumber\\
\end{eqnarray}
Here $\Delta=\varepsilon_{0}-\lambda^{2}/2$ is the polaronic
shift.

Then, the linearized force $F_{q}(t)$ in the right-hand side of
the Eq.(\ref{29}) is represented as the sum of two contributions:
$ F_{q}=F_{0}+F_{1}$, where
\begin{eqnarray}
 \label{A6}
F_{0}=\lambda \langle
\hat{c}_{0}^{+}(t)\hat{c}_{0}(t)\rangle\nonumber\\
-\sum_{k,j=-/L,+/R}j\lambda_{t}\bar{t}_{0j}\langle
\hat{Q}_{j}^{+}(t)\hat{a}_{kj}^{+}(0)e^{i(\varepsilon_{k}-\mu_{j})t}\hat{c}_{0}(t)\rangle\nonumber\\
-\sum_{k,j=-/L,+/R}j\lambda_{t}\bar{t}_{0j}\langle\hat{c}_{0}^{+}(t)\hat{a}_{kj}(0)e^{-i(\varepsilon_{k}-\mu_{j})t}
\hat{Q}_{j}(t)\rangle\,
\nonumber\\
\end{eqnarray}
and
\begin{eqnarray} \label{A7}
\nonumber
 F_{1}=-\lambda_{t}\bar{x}(t)\sum_{k,j=L,R}\bar{t}_{0j}\langle \hat{Q}_{j}^{+}(t)\hat{a}_{kj}^{+}(0)e^{i(\varepsilon_{k}-\mu_{j})t}\hat{c}_{0}(t)\rangle\nonumber\\
 -\lambda_{t}\bar{x}(t)\sum_{k,j=L,R}\bar{t}_{0j}\langle \hat{c}_{0}^{+}(t)\hat{a}_{kj}(0)e^{-i(\varepsilon_{k}-\mu_{j})t} \hat{Q}_{j}(t)\rangle \nonumber\\
 - \sum_{k,j=-/L,+/R}j\lambda_{t}\bar{t}_{0j}\langle \hat{Q}_{j}^{+}(t)\hat{a}_{kj}^{+}(0)e^{i(\varepsilon_{k}-\mu_{j})t}\hat{c}_{1}(t)\rangle\nonumber\\
 -\sum_{k,j=-/L,+/R}j\lambda_{t}\bar{t}_{0j}\langle \hat{c}_{1}^{+}(t)\hat{a}_{kj}(0)e^{-i(\varepsilon_{k}-\mu_{j})t} \hat{Q}_{j}(t)\rangle\nonumber\\
 +\lambda\left[\langle \hat{c}_{1}^{+}(t)\hat{c}_{0}(t)\rangle+\langle
 \hat{c}_{0}^{+}(t)\hat{c}_{1}(t)\rangle\right]\,
\nonumber\\
\end{eqnarray}

To proceed further, we calculate the averages in
Eqs.(\ref{A6}),(\ref{A7}). In perturbation theory on the level
width $\Gamma_{0}$ the averages of boson and fermion operators
factorize
\begin{eqnarray}
\label{A8}
\langle\hat{Q}_{j}^{+}(t_{1})\hat{a}_{kj}^{+}(0)\hat{a}_{k'j}(0)\hat{Q}_{j}(t_{2})\rangle\nonumber\\
=\langle\hat{a}_{kj}^{+}(0)\hat{a}_{k'j}(0)\rangle\langle\hat{Q}_{j}^{+}(t_{1})\hat{Q}_{j}(t_{2})\rangle\,,
\end{eqnarray}
where
$\langle\hat{a}_{kj}^{+}(0)\hat{a}_{k'j}(0)\rangle=f_{j}(\varepsilon_{k})\delta_{kk'}$.
For noninteracting equilibrated electrons in the leads
$f_{j}(\varepsilon_{k})=[\exp(\beta(\varepsilon_{k}-\mu_{j}))+1]^{-1}$
is the Fermi-Dirac distribution function.

It is easy to calculate boson correlation function in
Eq.(\ref{A8}), assuming vibrons to be at equilibrium at
temperature $T$ (this is a plausible assumption for a weak
tunneling regime we are dealing with). The standard calculations
(see Ref.\cite{LMcK}) results in
\begin{equation} \label{A11}
 \langle\hat{Q}_{j}^{+}(t_{1})\hat{Q}_{j}(t_{2})\rangle=\sum_{l=-\infty}^{+\infty}F_{lj}(\beta)\exp(il(t_{1}-t_{2}))\,,
\end{equation}
where $F_{lj}(\beta)$ is defined in the main text (see
Eq.(\ref{34})).

Now, by substituting Eq.(\ref{A11}) into the
Eqs.(\ref{A6}),(\ref{A7}) and taking all time integrals in the
limit $t\gg \tau_{0}$, we obtain the right-hand side of
Eq.(\ref{29}) (i.e.$F_{q}$) as a linear functional of
$\bar{x}(t)$. In this case the "force" term $F_{0}$ does not
depend on time
\begin{eqnarray}
\label{A14}
F_{0}=\lambda\langle\hat{c}_{0}^{+}(t)\hat{c}_{0}(t)\rangle=\frac{\lambda}{2}\sum_{j=L,R}\sum_{l=-\infty}^{+\infty}F_{lj}(\beta)f_{j}(\Delta-l)\nonumber\\
=const\,
\nonumber\\
\end{eqnarray}
It determines the initial position of QD. The linear in
$\bar{x}(t)$ term $F_{1}$ takes the form of the right-hand side of
Eq.(\ref{31}).

\end{document}